\documentclass[lettersize,journal]{IEEEtran}
\usepackage{svg}
\usepackage[numbers]{natbib}
\usepackage{float}
\usepackage{hyperref}
\usepackage{pdfpages}
\usepackage{soul}
\usepackage{color}

\usepackage{stfloats}
\usepackage{balance}

\ifCLASSINFOpdf
\else

\fi

\usepackage{adjustbox,lipsum}
\usepackage{titlesec}
\usepackage{algorithm}
\usepackage{algpseudocode}
\usepackage{pdfpages}
\usepackage{amsmath}
\usepackage{booktabs}
\usepackage{multirow}
\usepackage{multicol}
\usepackage{tables}

\begin{document}
\title{Steganalysis of AI Models LSB Attacks}

\author{Daniel Gilkarov,~\IEEEmembership{Member,~IEEE,}
        Ran Dubin,~\IEEEmembership{Member,~IEEE,}
        % <-this % stops a space
\thanks{Ran. Dubin is from the Department of Computer Science, Ariel University Ariel, Israel. From Ariel Cyber Innovation Center, Ariel University Ariel, Israel e-mail: (rand@ariel.ac.il).}% <-this % stops a space
\thanks{Manuscript received ????; revised ????}}

\markboth{IEEE TRANSACTIONS ON INFORMATION FORENSICS AND SECURITY,}%
{Shell \MakeLowercase{\textit{et al.}}: A Sample Article Using IEEEtran.cls for IEEE Journals}


\IEEEpubid{\makebox[\columnwidth]{0000--0000/00\$00.00~\copyright{}2022 IEEE \hfill} \hspace{\columnsep}\makebox[\columnwidth]{ }}

% \author{\IEEEauthorblockN{Daniel Gilkarov}
% \IEEEauthorblockA{\textit{Department of Computer Science} \\ \textit{Ariel Cyber Innovation Center}\\ Ariel University, Israel\\
% daniel.gilkarov1@msmail.ariel.ac.il}\\
% \and
% \IEEEauthorblockN{Ran Dubin}
% \IEEEauthorblockA{\textit{Department of Computer Science} \\ \textit{Ariel Cyber Innovation Center}\\ Ariel University, Israel\\
% rand@ariel.ac.il
% }
% }

\maketitle

\begin{abstract}
Artificial intelligence has made significant progress in the last decade, leading to a rise in the popularity of model sharing. The model zoo ecosystem, a repository of pre-trained AI models, has advanced the AI open-source community and opened new avenues for cyber risks. Malicious attackers can exploit shared models to launch cyber-attacks. This work focuses on the steganalysis of injected malicious Least Significant Bit (LSB) steganography into AI models, and it is the first work focusing on AI model attacks. In response to this threat, this paper presents a steganalysis method specifically tailored to detect and mitigate malicious LSB steganography attacks based on supervised and unsupervised AI detection steganalysis methods. Our proposed technique aims to preserve the integrity of shared models, protect user trust, and maintain the momentum of open collaboration within the AI community. In this work, we propose 3 steganalysis methods and open source our code. We found that the success of the steganalysis depends on the LSB attack location. If the attacker decides to exploit the least significant bits in the LSB, the ability to detect the attacks is low. However, if the attack is in the most significant LSB bits, the attack can be detected with almost perfect accuracy.
\end{abstract}

\IEEEpeerreviewmaketitle

\section{Introduction}
\label{Introduction}
As the cyber threat landscape grows in complexity and sophistication, steganography has emerged as a particularly insidious technique malware creators employ. Steganography is the ancient art of hiding data within other non-secret data. In the past, steganography was used as a tool for covert communication and secret messages. Now, it has found a disturbing place in the realm of digital crime, offering cyber criminals an effective method of cloaking malicious code.

One of the most striking examples of this technique was seen in the Zeus Trojan variant \cite{zeus_stego}, which used steganographic methods to conceal its configuration file within a JPEG image. Recently, a new attack technique named GIFShell~\cite{Gifshell} presented a novel and potent threat, as it cleverly exploits various features of Microsoft Teams to serve as a command-and-control (C\&C) conduit for malware operations. This innovative strategy involves data exfiltration concealed within innocuous-looking GIFs, which successfully evades detection by Endpoint Detection and Response (EDR) systems and other network surveillance tools. A key prerequisite for the success of this attack is an already compromised system or user, which acts as the launchpad for this sophisticated cyber-attack.

The primary potency of steganographic malware is anchored in its capacity to seamlessly assimilate with ordinary, non-threatening data, which subsequently camouflages it from standard cybersecurity defenses. The extraordinary subterfuge of steganographically altered files lies in their perceived mundanity, frequently permitting them to evade detection by traditional intrusion detection systems, Anti-Virus (AV) software, and Endpoint Detection and Response (EDR) tools. As such, steganographic malware introduces an intricate challenge to the landscape of cyber defense strategies, calling for innovative measures to counteract its elusive nature.

The mounting use of steganography by cybercriminals underscores the urgent need for advanced detection techniques and robust defense strategies. In the arms race that characterizes our cybersecurity landscape, grasping and mitigating steganographic malware has become a priority for researchers and practitioners alike. Multiple papers have researched steganography attacks in images~\cite{subramanian2021image}, videos~\cite{liu2019video}, audio~\cite{alsabhany2020digital} and in Deep Learning models~\cite{wang2021evilmodel,wang2022evilmodel}.

Steganalysis is the countermeasure for steganography \cite{muralidharan2022infinite}. Steganalysis is the practice of detecting and revealing hidden information inside the data and potentially extracting the concealed information.

Steganalysis methods involve various techniques ranging from statistical analysis and machine learning to signal processing. The constant development and refinement of steganalysis strategies are crucial in combating the evolving steganographic methods utilized in cyber threats. The goal is not only to detect the presence of hidden information but also, if possible, to recover the original, concealed data. Previous steganalysis related works focused on images~\cite{muralidharan2022infinite}, video ~\cite{bouzegza2022comprehensive}, and audio~\cite{ren2022universal}. This paper is the first steganalysis work that detects Least Significant Bit (LSB) steganography attacks on deep learning models.

It's important to note that the success of steganalysis greatly depends on the level of complexity of the steganographic techniques utilized and the amount of hidden content. More advanced steganographic methods can be difficult to detect, highlighting the necessity for continuous advancements in steganalysis techniques. Image steganalysis is particularly complex, and a recent review~\cite{muralidharan2022infinite} indicates that even the most advanced image steganalysis methods have low detection rates, presenting numerous challenges for the future.

Artificial Intelligence (AI) model security is an important research subject. While malware steganography in images and other domains is well-known, the risk of AI model sharing and downloading is a new attack vector. Model Zoo platforms such as Hugging Face \cite{Huggingface}, PyTorch Hub \cite{pytorchhub}, and TensorFlow Hub \cite{tensorflowhub} offer a variety of free state-of-the-art pre-trained models for anyone to download and may be used to utilize malicious attacks by hackers. Today, model detection is focused on antivirus and serialization attacks~\cite{Huggingface}. However, steganography attacks are not addressed and handled. This study explores the intricacies of the LSB model weight steganographic attacks. It also aims to address the challenges encountered in identifying these attacks using steganalysis while underscoring the significance of such attacks.

% This paper presents methods for performing steganography in neural networks or, more generally, in anything that contains float data. These steganography techniques are ideally carried out inside neural networks that have finished their training since a neural network's training process changes the floats, and changes in the float values can corrupt a message hidden within floats. Using steganography, an attacker can manipulate bits in one or more model weights and embed any message that is made up of bits in these weights. In our digital world, a binary string represents everything within a computer, so anything digital can be embedded inside these weights. From the cyber-security perspective, this is potential for all sorts of malicious actions, for example, masking malware code or executable files inside a neural network shared online for it to be extracted later.

The key contributions of our research are enumerated below:

\begin{itemize}
\item Novelty: This study marks the first attempt to present steganalysis solutions for the Least Significant Bit (LSB) deep learning model weight attack, leveraging two distinct detection approaches: supervised and unsupervised learning.
\item Feature Engineering: Our work proposes using two types of features - Reconstruction Loss~\cite{schürholt2022hyperrepresentations} and Back Propagation (BP) processes~\cite{rumelhart2013backpropagation}.
\item Methodology Development: We have established a novel framework for creating a learning dataset specifically tailored for steganalysis.
\item Promotion of Reproducible Research: In an effort to further the development in this nascent field, we have made our steganography attack tools, steganalysis methods, feature sets, and the corresponding code publicly available \cite{gitrepo}.
\item Limitations: We conclude the success and drawbacks of our proposed methods and present the future points for improvements.
\end{itemize}

The remainder of this paper is structured as follows. Section \ref{Related Works} describes the related works. Section \ref{sec:background} provides background knowledge about float32 data. Section \ref{Methodology} presents the Methodology for embedding the malware. Section \ref{sec:dataset} presents our dataset creation, and Section \ref{sec:data_exploration} explores and illustrates the data challenges. Section \ref{Evaluation} summarizes our evaluation and results. Section \ref{Future Work} summarizes the future work and research limitations. Finally, conclusions are summarized in Section \ref{Conclusion}.

\section{Related Works}
\label{Related Works}

Steganography in digital data has been primarily centered around images \cite{image_steg}, audio \cite{audio_steg}, and video \cite{video_steg}. LSB steganography techniques exist in all these digital data types and are well-known, researched, and combated against through various proposed steganalysis methods \cite{boehm2014stegexpose}. Adversarial steganography methods evolved in response to steganalysis methods that counter them \cite{image_steg}. The LSB methods started from basic LSB replacement and progressed to more sophisticated and advanced methods (Pattern-Based LSB \cite{image_lsb_matching}, etc.). AI model LSB steganography works \cite{wang2021evilmodel, wang2022evilmodel, stegonet} were published with the same general methodology seen in LSB steganography of other data types - where the LSBs in data carriers are manipulated and replaced with an embedded hidden message while minimizing the noticeable effect on the carrier. In AI models, the LSB steganography is performed in the model parameters (weights, biases). 
AI model steganalysis is an unexplored research domain, and as far as we know, this is the first work that explores steganalysis in AI models.

\section{Background}
\label{sec:background}
\subsection{Float32}
\label{sec:float32}
Float32 numbers (IEEE 754 standard) \cite{wiki:float32} take up 32 bits in memory. These 32 bits are segmented into three parts - 1/8/23 for sign/exponent/mantissa.

Given a 32 bits long string (1's and 0's) \[B=b_{32}b_{31}b_{30}b_{29} \cdots b_{3}b_{2}b_{1}\]
The sign bit is $b_{32}$. The exponent bits are $b_{31} \cdots b_{24}$, denoted $E$. Finally, the mantissa bits are $b_{23} \cdots b_{1}$. The decimal value is calculated by 
\begin{equation}
    (-1)^{b_{32}} \cdot 2^{E-127} \cdot (1+\sum_{i=1}^{23}b_{24-i}2^{-i})
\end{equation}
In the following paragraph, we supply a general analysis of float32 by analyzing the different parts. The aim is to show why the least significant bits are called the least significant bits.
\begin{equation}
    (-1)^{b_{32}} \in \{-1, 1\}
\end{equation}
\begin{equation}
    \label{eq:exponent}
    E=(b_{31} \cdots b_{24})_2 \in \{0, 1,...,255\} 
\end{equation}
\[\Downarrow\]
\begin{equation}
    \label{eq:exponent_value}
    2^{E-127} \in \{2^{-126}, ..., 2^{127}\}
\end{equation}
\begin{equation}
     \label{eq:mantissa_value}
    1+\sum_{i=1}^{23}b_{24-i}2^{-i} \subset [1, 2-2^{-23}] \subset [1, 2) 
\end{equation}
%\in \{1, 1+2^{-23}, ..., 2-2^{-23}\}
Equation \ref{eq:exponent_value} holds the value the exponent bits are responsible for in the float, and equation \ref{eq:mantissa_value} holds the value the mantissa bits dictate in the float value. The exponent's bits are therefore responsible for exponential changes in the float value, as the name suggests. On the other hand, the mantissa bits deal with small changes in the value in the range $[1,2)$ as seen in equation \ref{eq:mantissa_value}. The term "least significant bits" usually refers to the mantissa region since these bits only change the float value slightly. Through the same logic, the term "most significant bits" is usually synonymous with the exponent region.
These terms can take on an even more general perspective. Through careful analysis, one can come to the conclusion that a bit $b_i$ is always more influential in terms of float value than $b_j$ whenever $ 31\ge i>j\ge1$ - \textit{note this terminology of LSB and MSB excludes the sign bit since it has a special and different role in the float in contrast with the exponent and mantissa bits which only affect absolute value.}
% \hl{we should add this as one of the contribution formulations of LSB attack sensitivity?}
The analysis can be divided into 3 cases: 
\begin{enumerate}
    \item $i,j$ are both exponent bits
    \item $i$ is an exponent bit, $j$ is a mantissa bit
    \item $i,j$ are both mantissa bits
\end{enumerate}
Case 1: Looking at the definition of $E$ in equation \ref{eq:exponent}, it is clear $b_i$ will have more influence since $E$ is an unsigned binary number; $b_i$ being 0 or 1 changes the value of $E-127$ in equation \ref{eq:exponent_value} by $2^i$ as opposed to $b_j$ changing the value by $2^j$, which is smaller.

\noindent Case 2 is justified through our initial analysis comparing the exponent value with the mantissa value.

\noindent Case 3 can be justified through the same sort of analysis done in Case 1.

This more general point of view on the terms LSB and MSB points to the fact that even within the mantissa (the least significant bits in the float), there are MSBs and LSBs. So we can refer to different regions in the float32 structure through the scope of MSBs and LSBs, for example, the MSBs of the LSBs, the MSBs of the MSBs, etc. This perspective of LSB and MSB is local and you can say that in float32 - the 32nd bit ($b_1$) is the least significant bit in the float.
% and at the same time it's the 32nd most significant bit - but of course, this doesn't convey any meaning.

We will now define the terms for the rest of the article. We denote the X least significant bits as follows:
\begin{equation}
    XLSB := b_{X} \cdots b_1
\end{equation}
% \hl{not clear: "placeholder, in usage an integer" too complicated simplify the discussion}
We abbreviate it as XLSB, for example, 13-LSB refers to $b_{13} \cdots b_1$. In this setting, the appropriate values for X are integers in the range $[1,23]$. This confines XLSB to the mantissa region.
In the same fashion, we denote the X most significant bits in the following manner: 
\begin{equation}
    XMSB := b_{31} \cdots b_{32-X}
\end{equation}
In this setting, the appropriate values for X are integers in the range $[1,8]$, which confines XMSB to the exponent region.

\section{Methodology}
\label{Methodology}
In this section, we will illustrate the architecture and evaluation methodology.
\begin{figure*}[!h]
\centering
\includegraphics[width=\textwidth]{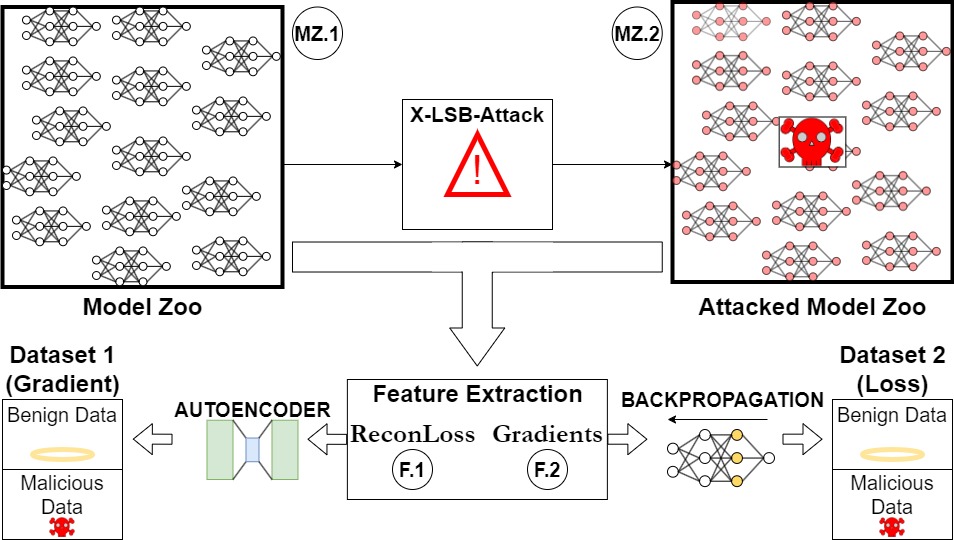}
\caption{Illustration of model LSB steganalysis architecture. A model zoo (MZ.1) and a model zoo for which every model in (MZ.1) was attacked with X-LSB-Attack (MZ.2) go through feature extraction - for every feature (F.1), (F.2), and so on... a dataset with the feature extracted from every model (MZ.1) and (MZ.2) is created (1 feature per model). Section \ref{sec:lsb_attack} details X-LSB-Attack, section \ref{sec:features} details the features and how they are extracted, and section \ref{sec:dataset} describes the dataset creation process.}
\label{fig:architecture}
\end{figure*}
\subsection{Workflow}
\label{Workflow}

% The process of researching LSB steganography on models used in the article involves two main parts: simulating an LSB attack and identifying valuable features for use in supervised/unsupervised learning techniques.
The article outlines two main steps for researching LSB steganography: simulating an LSB attack and identifying valuable features for use in supervised/unsupervised learning.

\subsubsection{LSB Attack}
\label{sec:lsb_attack}

\begin{figure*}[!h]
\centering
\includegraphics[width=\textwidth]{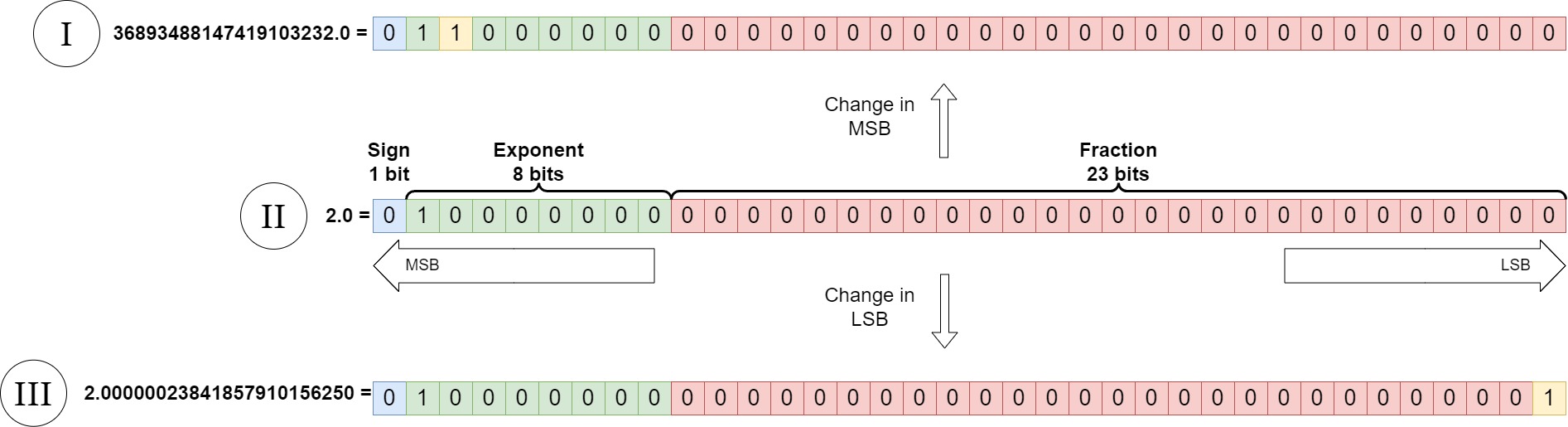}
\caption{Illustration of float32 structure and effect of an LSB vs. an MSB. Looking at (II) as a base, in (I), the 2nd MSB is changed and colored yellow. In (III), the 1st LSB is changed and colored yellow. The change between (III) and (II) is very small - about $2\times10e-7$. The change between (I) and (II) is very big - about $3.6893488e+19$. This is why the most significant bits are called that - they affect the float value the most and vice versa for the least significant bits. See section \ref{sec:float32} for an explanation of the composition of float32.}
\label{fig:msb_vs_lsb_change}
\end{figure*}

This section will discuss how the LSB attack targets a model's weights, typically of the float32 or float64 data type. Our methodology can be applied to different precision-point data. To hide data, changing bits inside the least significant bits (LSBs) of the float's fraction is wiser. Neural networks depend on the delicate precision of the float's fraction and usually keep the weights in the [0,1] range. Introducing a weight with changes in the exponent would likely hinder the model's performance. Our goal is to embed messages without the carrier's user noticing. Figure. \ref{fig:msb_vs_lsb_change} illustrates the effect of changing an LSB vs. changing an MSB.

\begin{figure*}[!h]
\centering
\includegraphics[width=\textwidth]{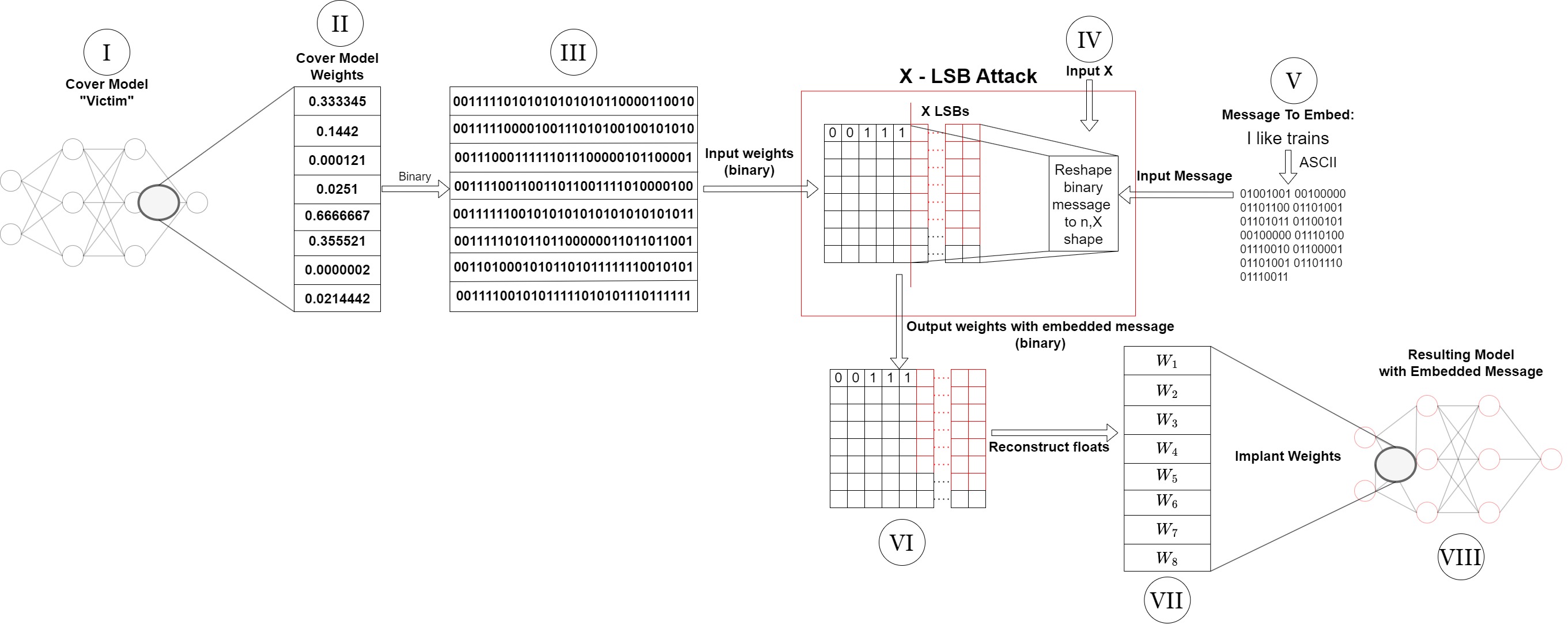}
\caption{Illustration of X-LSB attack - The weights (II) from a cover model (I) are extracted and stacked to form a column vector. Every float is transformed into a binary representation (III), an integer $1 \le X \le 23$ (IV), and a binary string (V) which are inputted into X-LSB-Attack. The attack embeds (V) into XLSB of the weights in (III) starting from the first weight. The resulting output is (VI), a binary matrix like (III) where (V) is embedded in the XLSB regions. (VII) is a float column vector constructed from each row of (VI), and then a model with the same architecture of (I) is initialized with the weights in (VII) - this is the resulting "attacked" model. See section \ref{sec:lsb_attack} for an in-depth explanation of the X-LSB-Attack depicted in this figure.}
\label{fig:x_lsb_attack}
\end{figure*}

We attack the model by embedding a malicious binary string inside the least significant bits of the model's weights. We name this attack \textbf{X - LSB Attack}. The attack takes as input the amount of bits to be attacked (denoted $X$). Since we're only talking about manipulating the bits of the mantissa,  we require $ 1 \le X \le 23 $. The embedding procedure essentially takes a binary string s, breaks it up into X-long segments (and a remainder segment when the length of s doesn't divide by X), and injects these segments one after another into the last X bits of the float weights of the model. See algorithm \ref{alg:x_lsb_attack} for pseudo-code and Figure \ref{fig:x_lsb_attack} for illustration.  
\begin{algorithm}
\caption{X - LSB attack}\label{alg:x_lsb_attack}
\hspace*{\algorithmicindent} \textbf{Input:} Integer $1 \le X \le 23$, AI model M, binary string s \\
\hspace*{\algorithmicindent} \textbf{Output:} AI model $\widehat{M}$ with s embedded in its weights
\begin{algorithmic}[1]
\State let $n_s:=$ length of binary string s
\State let $W:=$ flattened vector of the weights in M's layers' of shape $(n_W, 1)$
\If{$n_s > n_W \cdot X$}
    \State s cannot fully fit inside M, terminate.
\EndIf
\State Calculate $BW$, a binary matrix where every float in $W$ is expanded to a 32-length row vector of bits ($BW$'s shape is $(n_W, 32)$)
\State set $q:=\lfloor n_s / X \rfloor$, $r:=n_s mod X$
\State set $BW_{block}:=$ first $q \cdot X$ bits of s reshaped to square matrix of shape $(q, X)$
\State set $BW_{remainder}:=$ last $r$ bits of s
\State paste $BW_{block}$ into right upper corner of $BW$
\State paste $BW_{remainder}$ into $q+1$st row of $BW$
\State calculate $\widehat{W}$ by reconstructing the floats from $BW$
\State \Return $\widehat{M}:=$ AI model of the same architecture as $M$ with $\widehat{W}$ as its weights
\end{algorithmic}
\end{algorithm}
%in research and isn't supposed to be practical,
We create a second version of the \textbf{X - LSB Attack} algorithm, called \textbf{X - LSB Attack - Fill}, which is used for simplicity in research. We use all available bits in the victim model weights in this version. If the message s fits more than once, we inject it repeatedly until all bits are used (and the last multiple of s can get cut off in the middle). If s does not fit even once in the available space, we use the first $n_W \cdot X$ bits of s.
% \hl{what is the motivation for this mode?}

\subsubsection{Features}
\label{sec:features}

\paragraph{Reconstruction Loss}
% An autoencoder can be used to learn hyper-representations of model zoos \cite{schürholt2022hyperrepresentations}.
% When trained on model zoos, autoencoders learn to some extent what a model from that zoo looks like, we take advantage of this property by extracting the reconstruction loss value to be used as a feature in classification.
% When the autoencoders train, their aim is to minimize the loss value (as is traditional in learning models).
% The motivation behind this is to learn autoencoder to learn the Deep Learning model expected loss values as the baseline and detect the attack as Out Of Distribution (OOD) or anomaly - if a model with modified weights is inputted the loss value will be higher and that's something that may be detectable.
An autoencoder can be used to learn hyper-representations of model zoos \cite{schürholt2022hyperrepresentations}. When trained on a set of models of the same architecture, autoencoders learn what a model from that set looks like. We can use this property by extracting the reconstruction loss value to be used as a feature in classification. Autoencoders aim to minimize the loss value during training. The motivation behind this is to learn the Deep Learning model's expected loss values as the baseline and detect attacks as Out Of Distribution (OOD) or anomalies. If a model with modified weights is inputted, the loss value will be higher and detectable.
%The thought behind looking at the loss value is that because the autoencoder trained on benign models - if a model with modified weights is inputted the loss value will be higher and that's something that may be detectable.

We formulate the process of calculating the loss value as follows:
%The process of calculating the loss value is as follows:
Given an input model $M$
using an autoencoder $AE$ that trained on some model zoo:
\begin{equation}
    \begin{gathered}
        W:=W(M)\\
        \widehat{W}=AE(W)\\
        loss=loss_{AE}(W,\widehat{W})
    \end{gathered}
\end{equation}
where $W(M)$ denotes the weights of $M$, $AE(W)$ means inputting $W$ into $AE$ and $loss_{AE}$ means the loss function used in $AE$.

\paragraph{Backpropagation}

% When we try to identify unusual modifications in models, we search for signs of a "healthy" model. Backpropagation (BP)\hl{cite} is a common technique used in deep learning models where the weights are adjusted to become more efficient for the given task. Consequently, we identify a model as harmless by examining its BP process. The idea is that a model with its weights modified will regard a specific content differently (OOD) compared to benign models. Therefore, by using BP, we can emphasize how differently the model is familiar and similar to the rest of the model BP outputs.

When identifying unusual modifications in models, we examine signs of a "healthy" model. Backpropagation (BP) \cite{rumelhart2013backpropagation} is a common technique used in deep learning models where weights are adjusted to become more efficient for the given task. We identify a model as benign by examining its BP process. If a model's weights are modified, it will regard a specific content differently (OOD) compared to benign models. Therefore, by using BP, we can emphasize how differently the model is familiar and similar to the rest of the model's BP outputs.

BP in a neural network model is the following procedure:
To perform BP - first forward propagation happens by inputting an input $x$ into the neural network (denoted by $f$), producing an output \[a = f(x)\] At each BP step, we step through the layers backwards from the output layer to the input layer, calculating the gradient at each layer. And for the last layer (layer $L-1$ in a neural network with $L$ layers) the gradient is calculated by 
\[\delta^L = \nabla_{a}\mathcal{J} \odot a_{L}(z_L)\]
where $\nabla_{a}\mathcal{J}$ is the partial derivative of the model loss function L with respect to the activation, $\odot$ is elementwise matrix multiplication, and $a(z_L)$ refers to the value output by the activation function in layer L on the output of the Lth layer.
The gradients of layers $L-1$ to $1$ are calculated by
\[\delta^{L-1} = (W^{L-1})^T\delta^{L} \odot a_{L-1}(z_{L-1})\]
% See figure \ref{fig:backpropogation_calculation_dnn} for an illustration of BP in a typical feed forward DNN.
% In BP, gradients are intermediary products used in the process of training a neural network. We find additional value in these gradients, using them as features in our steganalysis process.

% In terms of the forward propagation step that's required for calculating gradients through BP, a need for input arises. We chose to input data that's all zeros, in theory, the input shouldn't matter too much, because the thought behind examining the gradients is that a group of similar models (model zoo) will produce similar gradients while models that are different (in this case through alteration of the weights) will produce outlier gradients with respect to the gradients of the model zoo.

Gradients are intermediary products used in the process of training a neural network and can be used as features in steganalysis. To calculate gradients through BP, input data is needed. We input zeros as the input data since the similarity between the gradients produced by a group of similar models (model zoo) can be examined, while outlier gradients produced by different models can be detected.

\subsection{Steganalysis Methods}
\label{sec:steganalysis_methods}
This section provides a brief overview of the methods used in our experiments.
% \hlcyan{The goal is to provide various ways to deal with the LSB attack detection challenge.}
We can categorize our methods into supervised and unsupervised methods.

\subsubsection{Supervised Methods}
\label{sec:supervised_methods}
Classification using supervised methods involves two main parts - data and methods. Labeled data is usually a dataset of features extracted from benign models and a dataset of features extracted from attacked models. When picking attacked data we can look at the 23 different levels of attack severity - 1 to 23 - where the more LSBs attacked (attack affects more significant bits) - the easier the challenge. In our trials, we use 3 different popular supervised learning models:
\begin{itemize}
    \item RandomForest
    \item XGBoost
    \item HistXGBoost
\end{itemize}
These models are selected because they are well-known methods in the literature and are highly effective.

\subsubsection{Unsupervised Methods}
\label{sec:unsupervised_methods}
Classification using unsupervised methods is done by learning a rule from the benign data and then testing it against the benign and malicious data.
In our trials, we use statistical properties to define thresholds.
In an anomaly-detection fashion, our unsupervised method classifies samples as "malicious" if they pass the threshold.
A simple threshold is defined by taking the mean value of the benign training data plus a "spacer" value. We call this technique $MEAN+\varepsilon$. The spacer value is picked from a predefined range of values by checking which spacer value gives the best classification score on the benign training data.

\subsection{Parameter Choice}
\label{section:parameter_choice}
The main parameter to consider in X-LSB-Attack is X, and it controls the severity of the attack. A higher X value means a more severe and impactful attack, and a higher X value means a greater capacity for embedding malware, but at the same time, it means the attack is more likely to change the way the model performs in a noticeable way, so choosing X is a balancing act between capacity for malware embedding and avoiding noticeable degradation of the cover data. Choosing X values in the high value ranges $[21,23]$ means the whole mantissa (or most of it) of the weights might change in AI models with normalized weights in the ranges $[-1, 1]$ or $[0, 1]$ which are almost fully represented by the mantissa bits, and the attack might introduce radical changes to the weights. In AI models, their "trained" state and proficiency for a certain task is represented by the layers' weights, so radical overall changes to a trained AI model's weights can render it useless for the task it has trained on, which in itself is a very big drawback from an attacker's point of view - no one would use a model which doesn't work, and on top of that the model might raise the suspicion of the end-user/model repository maintainer, etc.

\subsection{Attacked Model Zoo}
\label{sec:attacked_model_zoo}
For research, it is helpful to look at model zoos that all had their models attacked the same way. This helps to study how the attack affects the models on a more global scale, allowing research to find shared properties among the "infected" population of models or find phenomena that happen in contrast with the "healthy" population of models. Attacked model zoos are affected by the way they're attacked. There's importance to the choice of parameters passed into X-LSB-Attack (See section \ref{section:parameter_choice}), with different parameters producing different attacks and therefore producing different attacked model zoos. For research and simplicity, X-LSB-Attack-Fill is used in experimentation.
We used malware obtained from MalwareBazaar for embedded malware. The choice of $s$ is not important. What's important is that the same string $s$ with the same amount of LSBs is used in the attack on all models in the model zoo.

The method for creating an attacked model zoo is as follows:\\
Given a model zoo, set $\widehat{X}$ as the value we use for $X$ and choose a binary string $\widehat{s}$ to embed in the models. The attacked model zoo is created by applying X-LSB-Attack-Fill with $\widehat{X}$ and $\widehat{s}$ as parameters on every model in the zoo.

\section{Dataset}
\label{sec:dataset}

Datasets of features extracted from models serve us in research to create steganalysis models for classifying models as benign or malicious. To create datasets, we require model zoos, for example, a model zoo of CNNs. Datasets contain extracted features from models in the model zoo. A single feature is extracted for each dataset, and different features detailed in \ref{sec:features} have different benefits. Datasets are created by taking a model zoo, attacking it as detailed in \ref{sec:attacked_model_zoo}, and then extracting a selected feature from every model in the benign model zoo and every model in the attacked model zoo. The datasets are labeled benign and malicious, which is useful in supervised learning, and in unsupervised and supervised techniques, it confirms whether a model labels malicious models correctly.
% The datasets are created by taking a model zoo, extracting the selected feature from every model in that zoo - then attacking that zoo with X-LSB-Attack-Fill by app

% Given a model zoo, denoted $MZ$ - a collection of $n$ models. Set some ordering of the models in $MZ$ = $\{m_1, m_2, ..., m_n\}$. A Attacked model zoo $X-\widehat{MZ}$ is created by applying X-LSB-Attack-Fill on every model in $MZ$:
% \[X-\widehat{MZ}=\{XLAF(m_1), ..., XLAF(m_n)\}\]

%In our LSB steganalysis perspective, we take interest mainly in the weights from the layers of these models. 

\section{Data Exploration}
\label{sec:data_exploration}

\begin{figure}[!h]
\centering
\includegraphics[width=\columnwidth]{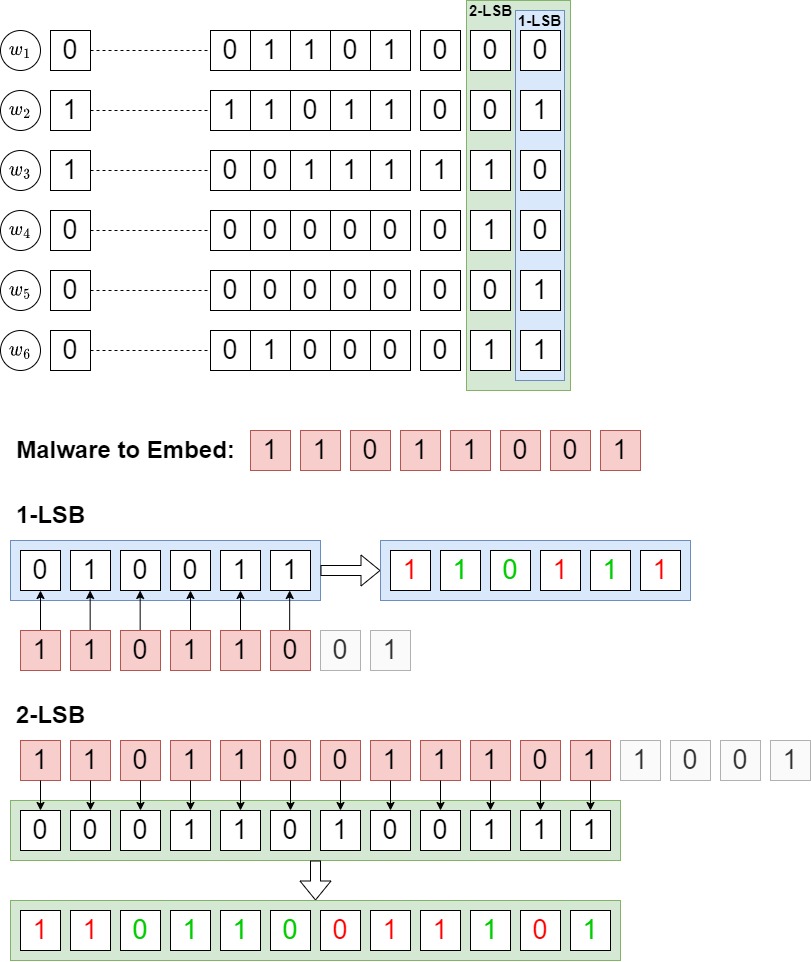}
\caption{Illustration of how the model weights change after embedding a binary malware string using X-LSB-Attack-Fill. In this example, the cover model has six weights - $w_1, ..., w_6$. The 1-LSB and 2-LSB regions of the weights are colored blue and green, respectively, and the malware we embed is colored red. Using X-LSB-Attack-Fill, the malware string is embedded repeatedly, with the excess (gray) discarded. The resulting 1-LSB and 2-LSB regions with the malware embedded in them are bits that changed in red color and bits that remained the same after the embedding in green. In the 1-LSB case, 3/6 bits remained unchanged after embedding; in the 2-LSB case, 6/12 bits remained unchanged. This illustrates the fact that when malware is embedded into the model parameters, it doesn't necessarily cause a change in the float values, and this is the main challenge for steganalysis since the aim is usually to find outliers by looking at the float values.}
\label{fig:data_exploration}
\end{figure}

This section aims to provide insight into how LSB steganography affects data from a high-level perspective and illustrates the challenges steganalysis methods have to overcome. When embedding binary malware inside the model weights, the LSBs in the floats are replaced with the binary data. There is a chance that bits that were replaced when embedding data stay the same (see Figure \ref{fig:data_exploration})

The fact that, in some cases, bits remain the same after embedding malware into the X-LSB region of the weights means that in some cases, some weights may remain completely unchanged after the embedding of malware. For example, in Figure \ref{fig:data_exploration}, when the 1-LSB was attacked $w_2, w_3, w_5$ remained unchanged, and when 2-LSB was attacked $w_2, w_3$ remained unchanged. Seeing this, we assume that fewer bits attacked means there is a higher chance of weights remaining unchanged. In the example in Figure \ref{fig:data_exploration}, for weights to remain the same when 1 bit was attacked, it only required 1 bit to match with the malware and for 2 bits attacked it required 2 bits to match. Note that in $w_5,w_6$ 1 out of 2 bits did match. Seeing this it is safe to assume the more bits we attack the smaller the chance for weights to stay the same, which means attacks with more bits are easier to detect.

We analyze the phenomenon noted above by looking at each attack model zoo and counting the number of weights that remained the same after embedding malware - the comparison is done per model. As expected, the more LSBs attacked, the fewer weights remained the same. We can even note that the mean amount decreases by half each time \#LSB is incremented. This seems to resemble the probability of the weights staying the same when the bits have a uniform independent probability of being 0 or 1, which is $1/2^{\#LSB}$.

\section{Evaluation}
\label{Evaluation}
In this section, the experimental results are laid out and analyzed. The experiments are carried out by applying classification techniques detailed in Section \ref{sec:steganalysis_methods} on features seen in Section \ref{sec:features}. The experiments are carried out using model zoos of CNN architectures
trained on 4 different datasets - (CIFAR10, MNIST, STL10, SVHN)   \cite{schürholt2022model}. The objectives for these experiments are: 
\begin{itemize}
    \item Provide the first successful steganalysis results in the AI LSB steganography field.
    \item Compare the success of different techniques and highlight weaknesses and strengths of different technique/feature combos.
    \item Measure success between different model zoos with different image data tasks.
\end{itemize}
Experiment 1 tests the unsupervised methods detailed in Section \ref{sec:unsupervised_methods}, and Experiment 2 tests the supervised methods detailed in Section \ref{sec:supervised_methods}.
In all the experiments, we used datasets created by the procedure detailed in \ref{sec:dataset}. Every model zoo produced datasets of 23 levels of attack severity (\#LSB attacked), and all experiments detail the classification scores for every model zoo with the same malware embedded into all the attacked model zoos. In all tables, Accuracy, Recall, Precision, and F1-Score are presented and abbreviated as A, R, P, F1. All results shown in tables and figures are the test results of the experiment. Notable results for every dataset in the experiments are written in bold text.

\subsection{Experiment 1 - Unsupervised Methods}
\begin{figure}[!h]
\centering
\includegraphics[width=\linewidth]{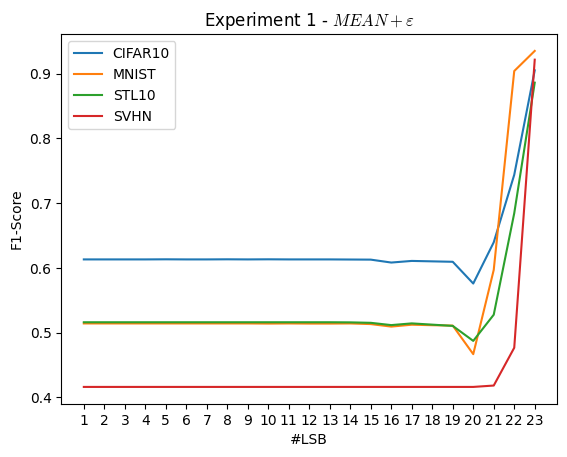}
\label{fig:exp1}
\end{figure}

Given a dataset, we use 70\% of the benign models' data in it for the training phase, and then the remaining 30\% of the benign models' data and all the malicious models' data are used for testing the results. 
In experiment 1, we apply the $MEAN+\varepsilon$ method detailed in Section \ref{sec:supervised_methods} on the losses dataset. We assume that malicious models will have a higher loss value, and this is one-dimensional data, so the $MEAN+\varepsilon$ is fitting for this case.
Table \ref{tab:exp1} details the classification scores.
Figure \ref{fig:exp1} shows a graph of F1-score vs. \#LSB modified.

% \begin{table*}[]
% \caption{Experiment 1 - mean + $\varepsilon$}
% \label{tab:exp1}
% \centering

% \expA
% \end{table*}

This method gives us better results than random guessing for $X\ge21$ in the CIFAR10, MNIST datasets, $X\ge22$ for SVHN, and in STL10, we only have success with 23 bits modified. The unsupervised methods have a clear advantage over the supervised methods - we only apply rules learned from the benign data. This means these models don't learn a specific attack and can generalize better for different attacks than supervised methods. The experiment results show that $MEAN+\varepsilon$ on STL10 had very minor success.

\subsection{Experiment 2 - Supervised Methods}
In the supervised learning experiments, we use 80\% of the benign and malicious data for training and the rest for testing. Each supervised experiment takes one dataset of features and trains a supervised classifier on that data. Along with the loss features and the gradient features, we also perform an experiment using the model weights as features themselves. In the same way with the extracted feature data, we have a model zoo, and the same zoo attacked as labeled data. In the results of our experiment, we abbreviated GradientBoosting to GB, HistogramGradientBoosting to HGB, and RandomForest to RF. All supervised experiments were conducted using the Scikit-Learn Python package.
% Table \ref{tab:exp2_reconloss} and Figure \ref{fig:exp2_reconloss} show the results from the experiment with loss features.
% Table \ref{tab:exp2_grads} and Figure \ref{fig:exp2_grads} show the results from the experiment with gradient features.
% Table \ref{tab:exp2_weights} and Figure \ref{fig:exp2_weights} show the results from the experiment with the model weights as features.
\subsubsection{Experiment 2.1 - Loss Features}
\noindent See Table \ref{tab:exp2_reconloss} and Figure \ref{fig:exp2_reconloss}.

The loss features are unique in the sense that they're one-dimensional data. The benefit of the loss features is that they provide a "higher"-level characterization of the model zoo. This method seems to fall short in classifying the CIFAR10 model zoo in comparison with the other model zoos. On MNIST, we get an F1-score of 63\% with GB for $X=21$. On STL10, we get an F1-score of 61\% with GB for $X=16$ and 59\% with GB for $X=17$. The best results in all datasets were achieved with the GB model.

\subsubsection{Experiment 2.2 - Gradient Features}
\noindent See Table \ref{tab:exp2_grads} and Figure \ref{fig:exp2_grads}.

Gradient features are multi-dimensional in contrast with the loss features. We see better results for the CIFAR10 dataset compared with the loss features - F1-score of 68\% using the RF model for $X=20$. On MNIST, we get an F1-score of 67\% using RF for $X=21$, which is also an improvement. On STL10, there wasn't success in the $X\in {16,17}$ region, but the results for $X=21$ using RF are better than in Experiment 2.1. On SVHN, the results are generally worse than in Experiment 2.1. The best results in all datasets were achieved with the RF model.

\subsubsection{Experiment 2.3 - Model Weights}
\noindent See Table \ref{tab:exp2_weights} and Figure \ref{fig:exp2_weights}.

By using weights for classification, we get very good results all using the HGB model. In all datasets, results for $x \ge 18$ are consistently high. The attack on the models attacks all the weights in all layers, leaving a global footprint across the weights. In practice, when embedding malware, there may be times when not all weights are changed. Another drawback of this method is that large models can hold a large amount of weight. This means the weights are very high-dimensional data, which takes a long time to train. While this method works better in this scenario, it is safe to assume that the previous methods are more robust and can handle different attacks, more local attacks that happen only in certain model regions.

\section{Future Work And Limitations}
\label{Future Work}

The future of AI model steganalysis is found in its first steps. When we started this research, we wanted to generalize the steganalysis detection for every model as a generic detection since AI models are diverse in their architecture and learning parameters. However, this direction could have been more successful. Therefore, we changed the methodology to focus on model architecture and creating training datasets for unsupervised and supervised steganalysis per architecture. 

Unsupervised learning was less effective compared to supervised learning. However, it has the advantage of being independent of the attack pattern knowledge. Like steganalysis in other domains, the ability to use steganography attacks inside the least significant LSB results in low detection but limits the size of the data that can be hidden inside the model. Future research can employ more advanced steganography methods that will be used to generalize the attack model training data better. Another direction is to use transfer learning from one model architecture to the other.

%Going forward with the AI Model attack research, there are different ways to further the steganalysis process.
%The steganography attack can be augmented, for example, only attacking certain model layers, distribution of the attack across the model, etc. From the steganalysis side, there is room for improvement in the unsupervised techniques. The threshold method can be refined to be less specific, allowing better results. Another unexplored option is trying a combination of the extracted features. Through careful analysis, selecting more meaningful features and discarding unimportant ones will be possible. In this article, all weights in the models were attacked. An interesting thing to analyze is the success of the methods on models that were partially attacked and also the success of the models that were trained in this article on partially attacked models. The generalizability of the models is a great asset.

%In this article, the steganalysis methods create models that learn model zoos of specific model architectures and tasks. This may be necessary as different architectures have different statistical properties, different amounts of weights, different distributions of layer types, etc.
% Future research can analyze the transfer learning ability of these steganalysis models.
\section{Conclusion}
\label{Conclusion}
This article introduces new steganalysis techniques to identify AI model steganography and safeguard publicly distributed models. Given the increasing popularity of AI model sharing, protecting models against attacks is a crucial problem. Our methods involve fine-tuning the models using various parameters and implementing a steganography model attack to create variations for supervised learning. We have introduced two different approaches: one is based on supervised and the other on unsupervised learning. We suggest three feature extraction methods: the first is centered around reconstruction loss, the second involves backpropagation, and the third involves learning the AI model weights. Our results show almost perfect detection in all methods when the attack targets the most significant LSB. However, we have found that sophisticated attackers who target their attacks on the least significant bits in the LSB can avoid detection, and we explain the complexity of defending AI models. Our open-source steganography and steganalysis framework encourages further improvements in this vast new field.

\section{Acknowledgement}
 This work was supported by the Ariel Cyber Innovation Center in conjunction with the Israel National Cyber Directorate in the Prime Minister's Office. This work is under US Provisional Patent Application No. 63/524,681.

\section{About the Authors}
% \noindent Daniel Gilkarov - Msc student in Ariel University.
% \noindent Ran Dubin - PhD, Lecturer, Department of Computer Science, Ariel Cyber Innovation Center
\vskip 0pt plus -1fil
\begin{IEEEbiography}
[{\includegraphics[width=1in,height=2in,keepaspectratio]{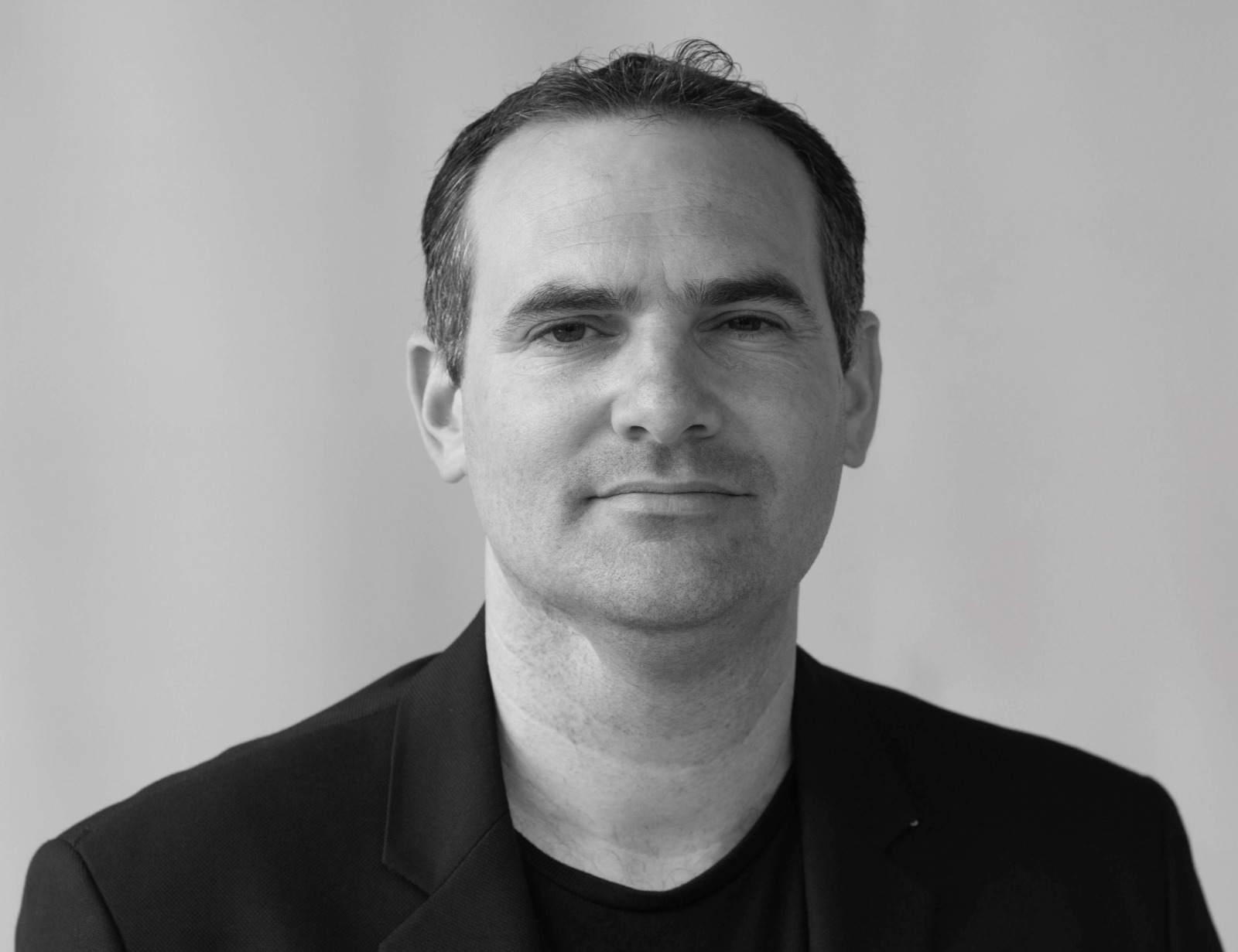}}]{Ran Dubin}
Ran Dubin received his B.Sc., M.Sc., and Ph.D. degrees from Ben-Gurion University, Beer Sheva, Israel, all in Communication Systems Engineering. He is currently a faculty member at the Computer Science Department at Ariel University, Israel. His research interests revolve around zero-trust cyber protection, malware disarms and reconstruction, encrypted network traffic detection, Deep Packet Inspection (DPI), bypassing AI, Natural Language Processing, and AI trust enhancements. 
\end{IEEEbiography}

\begin{IEEEbiography}
[{\includegraphics[width=1in,height=2in,clip,keepaspectratio]{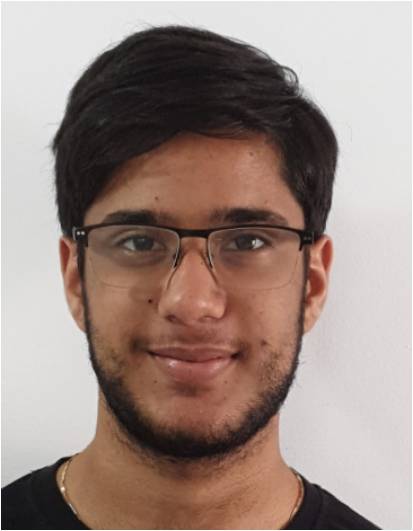}}]{Daniel Gilkarov}
Daniel Gilkarov received his B.Sc. in computer science from Ariel University and currently pursuing a M.Sc. degree in computer science. His research interests are deep learning and steganography. \end{IEEEbiography}

\bibliography{main}
% \balance

% \begin{table*}[h!]
% \caption{Amount of unchanged weights in attacked model zoos}
% \label{tab:data_exploration}
% \centering

% \dataexp

% \end{table*}

% \newpage
\onecolumn
\appendix

\begin{figure*}[!h]
\centering
\includegraphics[width=0.8\textwidth]{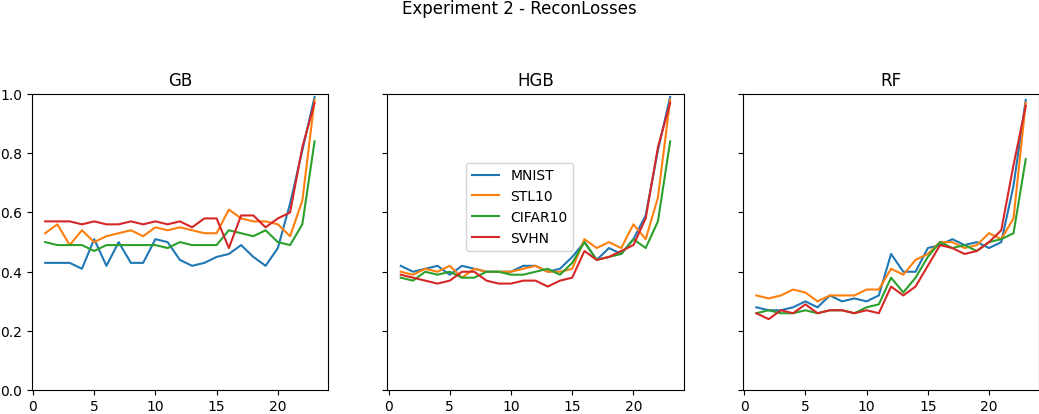}
\label{fig:exp2_reconloss}
\end{figure*}

\begin{figure*}[!h]
\centering
\includegraphics[width=0.8\textwidth]{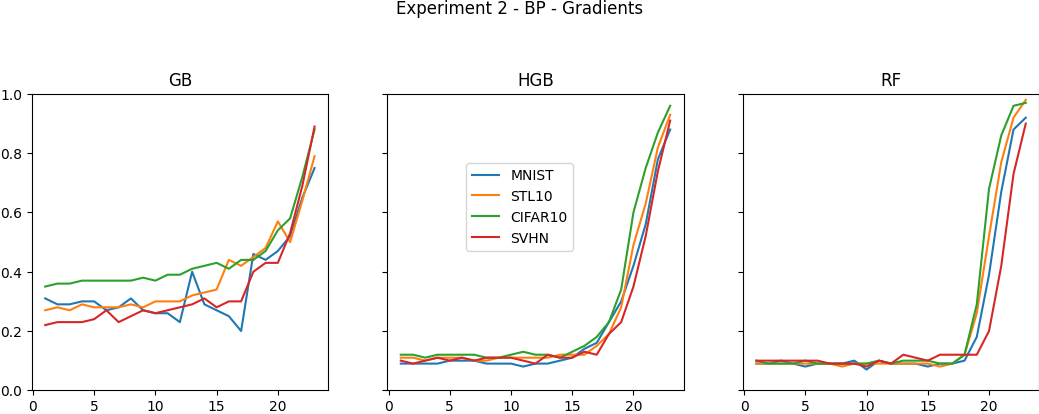}
\label{fig:exp2_grads}
\end{figure*}

\begin{figure*}[!h]
\centering
\includegraphics[width=0.8\textwidth]{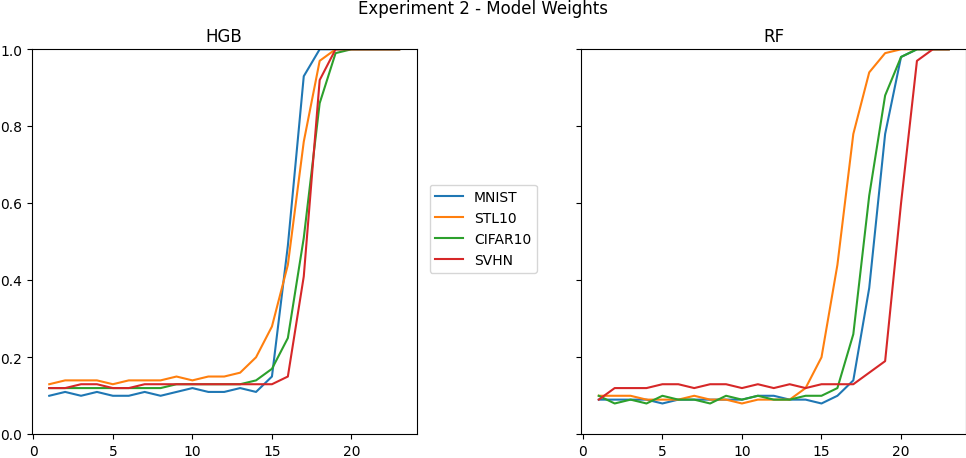}
\label{fig:exp2_weights}
\end{figure*}

\begin{table*}
\caption{Experiment 2 - ReconLoss}
\label{tab:exp2_reconloss}
\centering

\expBloss
\end{table*}

\begin{table*}
\caption{Experiment 2 - Gradients}
\label{tab:exp2_grads}
\centering

\expBgrads
\end{table*}

\begin{table*}[]
\caption{Experiment 1 - mean + $\varepsilon$}
\label{tab:exp1}
\centering

\begin{adjustbox}{height=100pt}
\expA
\end{adjustbox}

\vspace*{1 cm}

\caption{Experiment 2 - Weights}
\label{tab:exp2_weights}
\centering
\expBweights

\end{table*}

% \begin{table*}
% \caption{Experiment 2 - Weights}
% \label{tab:exp2_weights}
% \centering

% \expBweights
% \end{table*}

\end{document}